\documentclass[prl,twocolumn,showpacs,amsmath,amssymb,superscriptaddress]{revtex4}
\usepackage{dcolumn}
\usepackage{graphicx}
\usepackage{bm}
\begin{document}
\title{One dimensional gapless magnons in a single anisotropic ferromagnetic
nanolayer}
\author{A. Villares Ferrer}
\affiliation{Instituto de F\' \i sica, Universidade Estadual de Campinas,
13083-970, Campinas, SP, Brazil}
\author{P. F. Farinas}
\affiliation{Instituto de F\' \i sica, Universidade Federal do Rio de Janeiro,
21645-970, Rio de Janeiro, RJ, Brazil}
\author{A. O. Caldeira}
\affiliation{Instituto de F\' \i sica, Universidade Estadual de Campinas,
13083-970, Campinas, SP, Brazil}

\date{\today}

\begin{abstract}
Gapless magnons in a plane ferromagnet with normal axis anisotropy
are shown to exist besides the usual gapped modes that affect spin
dependent transport properties only above a finite temperature.
These magnons are one dimensional objects, in the sense that they
are localized inside the domain walls that form in the film. They
may play an essential role in the spin dependent scattering
processes even down to very low temperatures.
\end{abstract}
\maketitle

While charge and its transport play an essential role in carrying
and processing information, spin is the origin of magnetism and is
the basis to most of the storage devices in information-related
technologies. There is much interest in combining charge and spin
properties in order to expand the possibilities for applications
into the so called spintronics \cite{wolf}. This field, that has
started with the discovery of giant magnetoresistance (GMR) in
alternating Fe/Cr layers \cite{baib}, has rapidly become on focus
with the report of magnetic materials with high transition
temperatures ($T_C$) such as Ga$_{1-x}$Mn$_x$As ($T_C\sim 110$ K)
\cite{ohno} and Ti$_{1-x}$Co$_x$O$_2$ ($T_C\sim 300$ K)
\cite{ticoo}, for instance. With the upscaling of $T_C$ toward
room temperatures realizations of practical ideas ensue, pointing
in the direction of spin-based devices analogous to $p$-$n$
transistors \cite{vig}. Functioning principles of such devices
rely on the existence of domain walls (DW) in finite ferromagnetic
samples. The role of DW on transport of electrical current has
been shown in the early experiments on iron whyskers\cite{taylor}
and since then subjected to intense experimental and theoretical
investigation \cite{various}. The possibility that GMR
originates directly from DW may yield the fabrication of GMR
devices from single layered materials taking simultaneous
advantage of spin and charge degrees of freedom. With the
improvement of growing techniques, the possibilities of producing
2D ferromagnetic samples with DW of few tenths of \AA\phantom{,}
has become reality \cite{sample}.

\begin{figure}[t]
\centerline{
\includegraphics[scale=0.38]{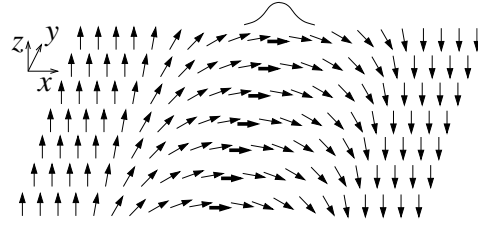}}
\caption{Domain wall in 2D showing the 1D
channel at its center where gapless magnons are
expected to propagate in the $y$ direction; the
amplitude of such a mode is depicted by the
bell-like curve shown on top.
\label{fig}}
\end{figure}

In this Letter we report on a magnetic excitation
that has not yet been accounted for the scattering of electrons by
DW. One-dimensional (1D) gapless magnons propagating inside a DW
of a single-layer ferromagnet are shown to exist in addition to
the usually known gapped magnons. Starting from an appropriate model
hamiltonian, we solve it for the continuum limit in 2D and find
that these modes are confined in the 1D DW, as
illustrated in Fig.\ref{fig}. Due to the absence of gap, the
inelastic coupling of these modes to electrons should
manifest at very low temperatures, as opposed to the usual magnon
scattering processes that become important only at temperatures
comparable to the anisotropy gap. We provide a theory to
include the
effect of the gapless magnons on conduction electrons.
We then present
results for the resistivity in a 2D system as a function
of the DW size, the electronic density, and the temperature
for realistic values of the parameters.

In order to model a system of independent electrons interacting
with local $1/2$ spins, we will use the effective hamiltonian
${\cal H}={\cal K}+{\cal H}_{H}+{\cal H}_I$, where
${\cal K}$ stands for the electron's kinetic energy operator
$-\hbar^2\nabla^2/2m$,
while ${\cal H}_H$ accounts for the interaction between local spins
(${\mathbf S}_i$), described by a 2D ferromagnetic Heisenberg spin
hamiltonian
\begin{equation}
{\cal H}_{H}=-J\sum_{i}{\mathbf S}_i \cdot{\mathbf S}_{i+1}
-J\Delta\sum_i({\rm S}_{i}^{z})^2, \label{dwe1}
\end{equation}
on a square lattice of constant $a$. Here $\Delta > 0$ gives the
strength of the on-site anisotropy and $J > 0$ is the exchange
amplitude. Interaction between electrons and the local spins is
considered through
\begin{equation}
\label{int}{\cal H}_I=-J_K a^2 \sum_i {\bm \sigma }\cdot {\mathbf
S}_i \delta ({\mathbf r}_i-{\mathbf r})
\end{equation}
where $J_K$ is the exchange integral and ${\bm \sigma}$ is the spin
of the electron.

First, we consider the static DW solution of ${\cal H}_H$ for a
$\pi $-domain. In the continuum the following
hydrodynamic equations are obtained for the magnetization,
\begin{equation}\label{eqte}
\frac{\partial \theta}{\partial t}=-j(\sin\theta \nabla^2 \phi
+2\cos\theta \nabla \theta \cdot \nabla \phi),
\end{equation}
and
\begin{equation}\label{eqfi}
\frac{\partial \phi}{\partial t}=j\left(
\frac{\nabla^2\theta}{\sin\theta} -\cos\theta (\nabla \phi)^2
-\frac{1}{\lambda^2}\cos\theta \right),
\end{equation}
with $j\equiv J S^2 a^2/\hbar$ and $\lambda \equiv a/\sqrt{2\Delta}$.
The fields $\theta $ and $\phi $ parametrize the magnetization
through the usual relation ${\bm S}({\mathbf r})=S(\sin{\theta }
\cos{\phi },\sin{\theta }\sin{\phi },\cos{\theta })$.
Such a hydrodynamic treatment suits
the low energy excitations studied in what follows, hence
the continuum limit is not only appropriate but becomes exact
for the long wavelength phenomena handled below.

It is possible to find a static solution for these coupled
equations in the form of a $\pi $ DW. Assuming $\phi = \bar{\phi
}$ is constant, $\theta = \bar{\theta }(x) $ (see Fig.\ref{fig}),
and using the boundary conditions $\lim_{x \rightarrow \mp\infty}
\cos\theta(x)=\pm 1$, the coupled equations can be easily
integrated to give $\bar{\bm S}({\mathbf r})=S [\hat{\bm x}\;
{\rm sech}(x/\lambda) -\hat{\bm z}\tanh (x/\lambda)]$
if we conveniently
choose $\bar{\phi }=0$.

Hence a $\pi $ DW like the one illustrated in Fig.\ref{fig} is
admitted as an equilibrium configuration in the 2D ferromagnetic
system described by ${\cal H}_H$. One sees that $\lambda $
measures the size of the spin-inversion region thus giving the
size of the DW. From the definition of $\lambda $ it is clear that
the DW extends throughout the sample unless the anisotropy is
different from zero. Strictly, the DW configuration is not the
absolute minimum of energy in our model, although it is a stable
solution. If one however takes the effect
of dipolar interactions, as in a finite sample,
the ground state becomes, in general, a DW distortion.
Yet, if the temperature is lowered, and as
long as the on-site anisotropy $\Delta >0$, one expects to
cross a transition temperature below which there are no
DW structures in the system. However,
in 2D one is able to induce DW's in a magnetic layer
up to a few thousand in a typical sample, even for
temperatures as low as a few K\cite{sample}. 
The use of this
particular solution as a starting
point is then justified, even at
the low temperatures for which the excitations studied ahead
are more important.

The DW solution for ${\cal H}_H$ as obtained here bares
resemblance to the solitons studied in the 1D Heisenberg chain
\cite{wri}. As we see below, this similarity
with the 1D chain for the equilibrium solution does not lead to
the same physics in 2D, the fluctuations about this solution are
given by the usual massive magnons {\it plus} the additional
gapless modes, only present in dimensions $> 1$.

We consider small deviations from equilibrium given by
$\theta=\bar{\theta}(x)+\xi({\mathbf r},t)$ and $\phi=\bar{\phi}+\eta
({\mathbf r},t)/\sin \bar{\theta }(x)$, where $\xi,\eta \ll 1$.
Substitution of these relations in Eqs.(\ref{eqte}) and
(\ref{eqfi}) yields
$
\partial^2 \xi ({\mathbf r},t)/\partial t^2=-j^2
[\nabla^2-V(x)]^2 \xi ({\mathbf r},t)
$,
where
$V(x)\equiv 
\left(1/\lambda^2\right)
\left[2\tanh^2{\left(
x/\lambda
\right)}-1\right]
$,
and an identical equation for $\eta ({\mathbf r},t)$.
Dispersion relations can then be
obtained from a standard choice of the solution in the form
$\xi ({\mathbf r},t)=b({\mathbf r})\exp{-i\omega} t$,
which leads to the following eigenvalue equation,
\begin{equation}
\left[\nabla^2-V(x)\right] b({\mathbf r})
= -\frac{\omega }{j} b({\mathbf r}). \label{modeqg}
\end{equation}

It is convenient to study the spectrum of the $x$
dependent solutions to understand why there must be a gap in 1D
but not in 2D. This is understood if we realize that
Eq.(\ref{modeqg}) yields the usual discrete eigenvalues associated
with the $x$ direction {\it and} a mode that is given by just
setting $\omega = 0$. Let us call this zero eigenvalue solution by
{\it zero-mode} for further reference. In 1D, the zero-mode is a
small {\it static} distortion from the original DW solution
localized near $x=0$ and with the functional form $\xi (x) \sim
{\rm sech}(x/\lambda )$, as it is seen by replacing
$x\rightarrow {\mathbf r}$ in Eq.(\ref{modeqg}) and integrating it
for $\omega = 0$. Since such a mode has no dynamics, the spectrum
of the ferromagnetic 1D anisotropic chain is gapped as it is well
known.
In particular, away from the DW $\lambda^2V(x)
\rightarrow 1$, and a solution is promptly obtained both in 1D and
2D, in the form of isotropic magnons with a massive spectrum given
by $\omega(k)=j( k^2+1/\lambda^2 )$ where $k$ is the wave vector.
This dispersion holds also in 2D with $k^2=k_x ^2 + k_y ^2$.

The key ingredient to distinguish 2D from 1D is that with the
additional degree of freedom, the zero-mode acquires dynamics and
becomes the gapless magnon that we are studying. We see this by
using a solution in the form $b({\mathbf r})=f(x)g(y)$ in
Eq.(\ref{modeqg}), which separates in
\begin{equation}\label{bnga}
\left[\frac{d^2}{dx^2}-V(x)\right]f(x)= \left(q^2 -\frac{\omega
}{j}\right)f(x),
\end{equation}
and 
\begin{equation}
g''(y)=-q^2 g(y).\label{bnga2}
\end{equation}
It is clear from these equations that
the zero-mode associated with the $x$ direction in Eq.(\ref{bnga})
is a {\it propagating spin wave}
with dispersion given by $\omega(q)= j q^2$.
In
2D this mode has a finite frequency and is localized about $x=0$,
as is seen from the solution $b({\mathbf r})\equiv {\rm
sech}(x/\lambda ) \exp{iqy}$ obtained from integrating
Eq.(\ref{bnga}) for $\omega(q)= j q^2$. An identical solution
holds for $\eta ({\mathbf r},t)$.
One can see that  $\xi $ is related to transverse ($\perp
\hat{\bm y}$), while $\eta $ to longitudinal
oscillations. The complete solution is given by
${\bm S}({\mathbf r},t)= \bar{\bm S}({\mathbf r})
+(\partial \bar{\bm S}({\mathbf r})/\partial
\bar{\theta})\xi ({\mathbf r},t) + \hat{\bm y}S
\;\eta ({\mathbf r},t)$,
where $\xi $ and $\eta $ are more generally linear combinations
of the $q$ and $\omega $ dependent solutions worked out
above. The gapless
dispersion corresponds to plane magnetization waves
propagating with very small amplitudes inside the 1D DW.

It should be pointed that these are not the first gapless
excitations reported in a 2D anisotropic chain. Gapless kinks that
could propagate under very restricted conditions have been
reported previously \cite{japas}, and they consist basically of
topological excitations inside the DW. Within our approximations,
Eq.(\ref{bnga2}) is linear
and dispersive and therefore will not sustain such localized kink-like
solutions which will ultimately decay into ordinary spin waves.
Gapless dispersions have also naturally appeared in solutions of
the XXZ hamiltonian in 3D,\cite{jafez} showing that this
is not a particularity of the 2D case we are treating.
Here it becomes clear the
necessity of dimensions $> 1$ for the existence of such wall magnons.

Consideration of
these modes for the transport of electrons in magnetic
materials appears to be absent from the up to date
literature, where magnons are overwhelmingly considered
in the analysis of measured data only at temperatures
suffciciently high to be comparable to the anisotropy gap,
whose typical values range from 50\phantom{x}K to 150\phantom{x}K,
depending on the
material used. Absence of gap in the spectrum
should make the contribution of
these objects relevant for the resistivity at lower
temperatures.

In what follows, it will be
seen that a transport theory for electrons interacting
with the gapless magnons can in fact be put forward without
complications and that the effects on the resistivity are expected
to be quite observable for temperatures much lower than
the anisotropy gap.

Hydrodynamics is quantized
by writing
\[
\widehat{\xi }[qy]
= \frac{a}{2\sqrt{L_y\lambda }}
{\rm sech} \left(\frac{x}{\lambda } \right)
\sum_q\left[\exp{(iqy)}a_q+h.c.\right],
\]
and $ \widehat{\eta }= \widehat{\xi }[qy-\pi /2]$.
This
ensures that the contribution
of the free magnons in Eq.(\ref{dwe1}) is written
as ${\cal H}_m= \sum_q
\hbar \omega_q a^{\dag}_q a_q$,
where $a^{\dag}_q$ is the (Heisenberg) bosonic operator that creates
a magnon with wave vector $q$,
$\hbar \omega_q = JS^2 a^2 q^2$, and $L_y$ is the width of the
sample.

The electron-magnon interaction
given by Eq.(\ref{int}), can be put
in the form
${\cal H}_I=
U_0(x) + U_1(x) +{\cal H}_{em}$,
where
$
U_0(x)= SJ_K\sigma_z \tanh (x/\lambda )$,
$
U_1(x)= SJ_K\sigma_x {\rm sech}(x/\lambda )$,
and
\[
\frac{{\cal H}_{em}}{SJ_K} = 
 \left[\sigma_x \tanh \left(\frac{x}{\lambda }\right)
+\sigma_z {\rm sech}\left(\frac{x}{\lambda }\right)
\right]\widehat{\xi }
-\sigma_y\widehat{\eta }.
\]
The term
given by $U(x)=U_0(x)+U_1(x)$ affects the electron's motion
only in the
$x$ direction and can be incorporated as an
effective potential in a purelly
electronic hamiltonian,
${\cal H}_e=-\hbar^2 \nabla^2/2m + U(x)$.
With these definitions, the full hamiltonian
${\cal K} + {\cal H}_H +{\cal H}_I$ can be
rewritten as ${\cal H}_e+{\cal H}_m
+{\cal H}_{em}$, where effectively,
the two first operators
are single-particle hamiltonians while
the last one accounts for the interaction. 

As a result of its interaction with the
DW, the electron experiences an effective potential
that combines a spin-conserving term, $U_0(x)\propto \sigma_z$
and a spin-flipping term $U_1(x)\propto \sigma_x$. The former
has the approximate form of a smooth step barrier centered
at the DW while the
latter exists only within the DW. From this structure of $U(x)$ 
it is seen that spin-up electrons tend to be blocked from left to
right while the opposite is true for the spin-down electrons. This
is expected since the energy of one spin rises when it reaches the
region where the magnetization is opposed to it. The DW tends to
work then as a spin diode, since it will conduct one species more
favorably in one direction.

Interaction with
magnons, given by ${\cal H}_{em}$, may be viewed as a complex potential
in the $x$ direction, that besides flipping the spin of the
electrons, has also a dissipative character. Spin-flip makes transport
favorable since the electron can travel into the opposite-spin
region with an energy gain. However, the momentum lost by emitting a magnon
causes the electrons to loose energy in the
process, and this is a temperature dependent component. What we
expect, on a qualitative basis, is that these magnons will lower
the efficiency of such a diode as a spin filter, having the same
effect on the transport electrons as the usual hot magnons, but
showing their effect at much lower temperatures.

To calculate the conductivity due to such an interaction
we start with the Boltzmann transport equation, which may
be written in a relaxation time approximation by assuming that
the change in the electrons' distribution function is 
given by $\delta f^\sigma ({\mathbf k})
=C(k){\mathbf k}\cdot {\mathbf E}$, where ${\mathbf E}$
is an in-plane uniform
electric field
perpendicular to the DW. To leading order, the collisions with
the DW affect only the scatering ampliudes
${\cal W}_{{\mathbf k}{\mathbf k}'}^\sigma
\equiv W_{{\mathbf k}{\mathbf k}'}^{\sigma \sigma}
+W_{{\mathbf k}{\mathbf k}'}^{\sigma -\sigma}$,
where $W_{{\mathbf k}{\mathbf k}'}^{\sigma \sigma'}$
is the probability per unit time for an electron in an
initial state $|{\mathbf k}\sigma \rangle $ to be scattered
to $|{\mathbf k}'\sigma' \rangle $. With these assumptions,
the relaxation time obtained is
\begin{equation}
\frac{1}{\tau^\sigma ({\mathbf k})}=\frac{\cal A}{4\pi^2}
\int {\cal W}_{{\mathbf k}{\mathbf k}'}^\sigma
\left[ 1-\frac{{\mathbf k}\cdot {\mathbf k}'}{kk'}
-\tan{\theta }\frac{\left|
{\mathbf k}\times {\mathbf k}'\right|}{kk'}
\right]d{\mathbf k}', \label{relaxtime}
\end{equation}
where ${\cal A}$ is the area of the sample and $\theta $ is
the angle between ${\mathbf k}$ and ${\mathbf E}$.
The crossed term that appears on the right
side of Eq.(\ref{relaxtime}) is a particularity of the
2D system, it is absent in both 1 and 3 dimensions.
It does not modify the relaxation time considerably though
since, as it is known, the collisions that
mostly affect the transport current consist
of nearly backward processes (${\mathbf k}\sim -{\mathbf k}'$) with
exchange of momentum $\sim 2k_F$.
The conductivity is readily calculated from $\tau^\sigma ({\mathbf k})$,
\begin{equation}
\rho^{-1} = \frac{e^2\hbar^2}{m^2}\sum_\sigma \int k^2\cos^2{\theta }
\tau^\sigma ({\mathbf k})\delta (E_{\mathbf k}^\sigma - E_F^\sigma ).
\label{rho}
\end{equation}

The scattering
amplitudes are calculated
by using the exact states of ${\cal K}+U_0(x)+ {\cal H}_m$, which is the
only part in ${\cal H}$ that does not vanish far from the
DW. If we denote such states by $|{\mathbf k}\sigma n_q\rangle $,
where $n_q$
indicates the magnon state in occupation-number formalism, then
the amplitudes will be given by
$W_{{\mathbf k}{\mathbf k}'}^{\sigma \sigma'}=2\pi N/\hbar
\sum_{qq'}
|\langle {\mathbf k}\sigma n_q |V^{\sigma \sigma' }|{\mathbf k}'\sigma'
n_{q'}\rangle |^2\delta_{\cal Q}$,
where $N$ is the number of 
domain walls and $\delta_{\cal Q}$ constrains the transitions within the
ones that conserve total energy and momentum. Here,
$V^{\sigma -\sigma }=V_x+V_y$ and $V^{\sigma \sigma }=V_z$, while
$V_i$ is the sum of all terms in $U_1(x)+{\cal H}_{em}$ that are
$\propto \sigma_i$. As stated, the detailed solutions constitute text book
material, but are nonetheless sufficiently awkward not to
be shown here\cite{fullpaper}. The solutions are used to numerically
handle Eqs.(\ref{relaxtime}) and (\ref{rho}) in obtaining the resistivity
$\rho $.
\begin{figure}[t]
\centerline{
\includegraphics[scale=0.30]{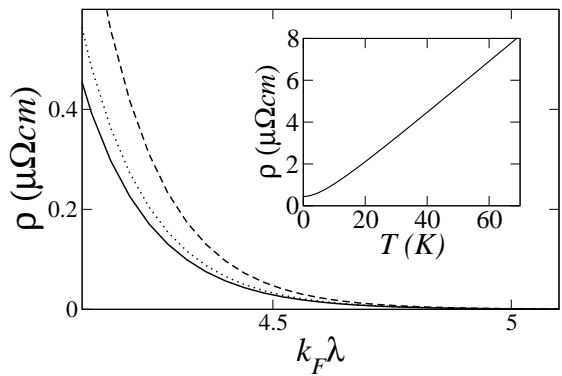}}
\caption{Resistivity in the $x$ direction as a function of
$k_F\lambda$ where $\lambda $ is the size of the DW. The
continuous line corresponds to the elastic case (no magnons),
while the dotted and dashed lines were computed including the
electron-magnon interaction for T=10 K and 25 K respectively.
The anisotropy gap used to obtain these curves was about 116 K (10
meV) and in all cases $J_K=0.5$ eV. The inset shows
the temperature dependence for $k_F\lambda \sim 4.2$. \label{fig4}}
\end{figure}

Fig.\ref{fig4} compares the resistivity along the $x$ direction of
a 2D sample for different values of the temperature as a function
of $k_{F}\lambda$ with and without including the gapless magnons.
The sample size has been taken as $L_x$ = 500 $\mu $m and $L_y$ =
50 $\mu $m, and $N\sim
3000$. Such a large number of DW are not expected
to be naturally
pesent in magnetic layers down to arbitrarily low
temperatures. However,
the interesting nanolayers for spintronics and
device applications should present DW at much lower
temperatures, and
such a number is achieved by experimental
techniques down to 2 K\cite{sample}.
If one considers the typical electronic densities, in the range from 10$^{15}$
to 10$^{20}$ cm$^{-2}$, in layers,
the DW sizes shown in Fig.\ref{fig4} range
from a few \AA\phantom{,} to several tenths of \AA, consistent with what is
also observed in experiments\cite{sample}.

One sees that an observable increase in the
resistivity results for most values of $k_F\lambda $, as
expected due to the localized character of the gapless magnons.
What is notable here, is that even at a temperature of 10 K the effect
can be quite observable, while the usual gapped magnons considered in
most of the literature would, in this case, affect the results
only for temperatures larger than about $116$ K (the anisotropy gap).
As a final remark, we would like to point
out that these effects can be experimentally checked by measuring
the magnetoresitance related to fields higher than the critical DW
field. This eliminates scattering processes as phonons or
non-magnetic impurities from the background, as it is known. If
the remaining resistivity shows a temperature dependence
consistent with a magnon bath for temperatures considerably lower
than the anisotropy gap, this will strongly suggest the presence
of the gapless magnons.

In closing, we have found that, contrary to the case in 1D and to
common belief, the magnetic excitation spectrum of a 2D
(single-layer) anisotropic ferromagnet should influence
charge transport at considerably low temperatures in spite of the
anisotropy. This is due to the existence of one dimensional
gapless magnons that can propagate in the domain walls formed as
the lowest energy configuration. We have provided a theory to describe
how electrons interact with both these magnons and the domain
wall. We have presented results
showing the effect of these magnons on the
transport of electrons through domain walls.

A.V.F. and A.O.C. wish to thank Funda\c c\~ao de Amparo \` a Pesquisa do
Estado de S\~ao Paulo (FAPESP) for financial support. P.F.F. and A.O.C.
kindly acknowledge partial support from Conselho Nacional de
Desenvolvimento Cient\'{\i}fico e Tecnol\'ogico (CNPq).

\vspace{0.4cm}
\hrule

\end{document}